\documentclass[preprint]{aastex63}  
\usepackage[utf8]{inputenc}
\usepackage{natbib}
\usepackage{hyperref}
\usepackage{graphics}
\usepackage{floatrow}

\usepackage[width=6.5in,height=9.0in,top=1.8in,left=1.3in,letterpaper]{geometry}

\date{2020 September 15}

\NewPageAfterKeywords
\begin{document}
\setcounter{page}{0}

\title{Community Challenges in the Era of Petabyte-Scale Sky Surveys}

\author[0000-0002-6702-7676]{Michael S.\ P.\ Kelley}
\altaffiliation{Phone: 301-405-3796; email: msk@astro.umd.edu}
\affiliation{University of Maryland, College Park, Maryland, USA}

\author[0000-0001-7225-9271]{Henry H.\ Hsieh}
\affiliation{Planetary Science Institute, Tucson, Arizona, USA}
\affiliation{Academia Sinica Institute of Astronomy and Astrophysics, Taipei, Taiwan}

\author[0000-0001-7335-1715]{Colin Orion Chandler}
\affiliation{Department of Astronomy and Planetary Science, Northern Arizona University, PO Box 6010, Flagstaff, AZ 86011, USA}

\author[0000-0002-1398-6302]{Siegfried Eggl}
\affiliation{DIRAC Institute and Department of Astronomy, University of Washington, Seattle, Washington, USA}
\affiliation{Vera C.\ Rubin Observatory, Tucson, Arizona, USA}

\author[0000-0003-0437-3296]{Timothy R.\ Holt}
\affiliation{Centre for Astrophysics, University of Southern Queensland, Queensland, Australia}
\affiliation{Department of Space Studies, Southwest Research Institute, Boulder, CO, USA}

\author[0000-0001-5916-0031]{Lynne Jones}
\affiliation{DIRAC Institute and Department of Astronomy, University of Washington, Seattle, Washington, USA}
\affiliation{Vera C.\ Rubin Observatory, Tucson, Arizona, USA}

\author[0000-0003-1996-9252]{Mario Juri\'c}
\affiliation{DIRAC Institute and Department of Astronomy, University of Washington, Seattle, Washington, USA}
\affiliation{Vera C.\ Rubin Observatory, Tucson, Arizona, USA}

\author[0000-0002-3818-7769]{Timothy A.\ Lister}
\affiliation{Las Cumbres Observatory, 6740 Cortona Drive Suite 102, Goleta, CA 93117, USA}

\author[0000-0001-5820-3925]{Joachim Moeyens}
\affiliation{DIRAC Institute and Department of Astronomy, University of Washington, Seattle, Washington, USA}
\affiliation{Vera C.\ Rubin Observatory, Tucson, Arizona, USA}

\author[0000-0001-5750-4953]{William J.\ Oldroyd}
\affiliation{Department of Astronomy and Planetary Science, Northern Arizona University, PO Box 6010, Flagstaff, AZ 86011, USA}

\author[0000-0003-1080-9770]{Darin Ragozzine}
\affiliation{Brigham Young University, Department of Physics and Astronomy, N283 ESC, Provo, UT 84602, USA}

\author[0000-0003-4580-3790]{David E. Trilling}
\affiliation{Department of Astronomy and Planetary Science, Northern Arizona University, PO Box 6010, Flagstaff, AZ 86011, USA}


\pagestyle{empty}

\clearpage
\pagestyle{plain}

\section{Introduction}

Large-scale surveys have had an enormous impact on many areas of astronomy and astrophysics, including solar system science.  Surveys such as the Catalina Sky Survey (CSS), Lincoln Near-Earth Asteroid Research (LINEAR) survey, and the Panoramic Survey Telescope and Rapid Response System (Pan-STARRS), have been prolific discoverers of small solar system objects, discovering hundreds of thousands of asteroids and hundreds of comets among them in the last twenty years\footnote{\url{https://cneos.jpl.nasa.gov/stats/site_all.html}}.  These discoveries lead to vital dynamical and physical information on small bodies, which in turn leads to improvements in our understanding of the origin and evolution of the solar system.
\looseness=-1

The next generation of large-scale surveys will continue these trends.  Those that are expected to have the largest impacts on solar system science include the Legacy Survey of Space and Time (LSST) to be conducted by the Vera C. Rubin Observatory (formerly known as the Large Synoptic Survey Telescope), the Near Earth Object Surveillance Mission (NEOSM), and NASA’s Nancy Grace Roman Space Telescope (formerly known as the Wide Field InfraRed Survey Telescope) mission \citep{schwamb2018_lsst_science_roadmap,holler2018_wfirst}.
\looseness=-1

In this white paper, we outline the challenges faced by the planetary science community in the era of next-generation large-scale astronomical surveys, and highlight needs that must be addressed in order for the community to maximize the quality and quantity of scientific output from archival, existing, and future surveys, while satisfying NASA's and NSF's programmatic goals.  Specifically, we ask the agencies to encourage efforts by the community to develop tools and infrastructure to facilitate research based on big surveys, allocate sufficient funding to support related work before and during such surveys, and develop the next-generation of data archives that will facilitate the analysis of large-scale data sets.
\looseness=-1

\section{Big Data Challenges for Solar System Science}

Petabyte-scale surveys should provide unprecedented scientific opportunities for solar system scientists, but also present a variety of technical, logistical, and sociological challenges.  In this white paper, we are primarily focused on the opportunities and challenges presented by the next generation of large-scale surveys mentioned above, although many of the issues we discuss are also applicable to the current and future management of planetary data from existing surveys such as Pan-STARRS, Asteroid Terrestrial-impact Last Alert System (ATLAS), Dark Energy Survey (DES), CSS, Transiting Exoplanet Survey Satellite (TESS), and the Zwicky Transient Facility (ZTF).
\looseness=-1

The big data threshold is the point at which common tools no longer work for the data volume, with additional consideration for complexity.  Most observational astronomers can analyze locally accessible data with software executed on their own CPUs.  However, in the next decade several observatories or sky surveys will have petabyte-scale data sets, all topped by LSST with a 200 PB data archive \citep{desai19}.  Astronomers will increasingly rely on online tools, services, or data platforms that can properly accommodate big data resources.  

The ultimate goal is to address the scientific priorities of the NSF and NASA, as they relate to planetary science.  Analyses of large data sets ultimately lead to the critical discoveries of unusual objects or rare phenomena, and to population-level characterizations. Population studies based on survey data have provided outcomes such as the Nice model \citep{gomes2005_nicemodel}, a paucity of small perihelion distance asteroids \citep{granvik2016_disruptions}, alignments in the trans-Neptunian population pointing to the possible existence of a giant planet in the distant outer solar system (see \citealt{2020tnss.book...79T} and references therein), and the identification of more than 100 asteroid collisional families \citep{2015aste.book..297N}.  Synoptic studies of individual objects through imaging surveys afford us a detailed understanding of specific processes and rare objects, such as the preservation of water ice in the main-asteroid belt \citep{hsieh2006_mbcs}, the discovery of the CO-dominated (i.e., water-depleted) comet C/2016 R2 \citep{mckay19}, and the first meteoritic samples of an astronomically observed asteroid, 2008 TC$_3$ \citep{jenniskens09}. All of these examples relied upon the previous and present generations of surveys.  Future scientific returns Rubin Observatory's first 10 years is summarized in a companion decadal white paper \citep{sssc-science-decadal2020}.
\looseness=-1

\section{Key Areas of Focus}
Within the context of a scientific investigation, planetary astronomers need to: (1) obtain appropriate project funding; (2) identify which data products are useful, and which pixels or database rows therein need to be analyzed; (3) access and retrieve that data; (4) analyze it; and, (5) consider follow-up studies.  These are generically stated, but important considerations arise from the magnitude of the data set volume, which affects methods and data systems.  Within this context, we address several key areas of focus in detail.
\looseness=-1

\subsection{Software Development}
The wide range of software development needs that should be addressed to maximize the scientific output of LSST data (and will likely be relevant for NEO Surveyor and Roman Space Telescope data as well) has been discussed in detail by \citet{hsieh2019_lsst_software}.  For instance, open-source and community-supported implementations of advanced discovery and detection algorithms are necessary to enable discovery of solar system objects at scale in survey and archival data sets. Examples that are currently in development include THOR \citep{THOR2019}, KBMOD \citep{KBMOD2019}, and HelioLinC \citep{Holman2018}.  Moreover, a survey-agnostic toolkit for rapid and basic asteroidal and cometary characterization (such as activity detection or taxonomic classification) could jump-start the scientific output of any telescope, survey, or data archive.  The planetary science community needs simple but robust, well-tested, open-source code-bases that are easily deployed on any large-scale data set.
\looseness=-1

The NASA Planetary Data Archiving, Restoration, and Tools (PDART) Program is a potentially important resource that must be adequately funded, and allow projects that support NSF-funded observatories.  Furthermore, projects must be allowed even before observatories are commissioned.  With paradigm-shifting facilities, such as the Rubin Observatory, community preparation is key to efficient data use during the survey.
\looseness=-1


\subsection{Follow-up Observations, Acquisition, and Management}

The present and near-future ground-based surveys are fundamentally time-domain observational systems. Those observatories that are or will be releasing near-real time transient alert streams (e.g., ZTF, Rubin Observatory, ATLAS) enable the solar system community to rapidly respond to transient phenomena, e.g., to characterize meter-scale NEAs or to study the gas in a cometary outburst. Connecting these alert streams to traditional observatories for follow-up screening or detailed study is the business of observation managers \citep{street2018_lsst_observing_management}. Observers could specify their trigger criteria, which would facilitate follow-up observation requests at other observatories, perhaps even automatically.  NEOExchange\footnote{\url{https://neoexchange.lco.global/}} is one such example \citep{lister2016}.  However, these tools still need development, especially integration with solar system databases (e.g., NASA's Planetary Data System (PDS) or the Lowell AstorbDB), and prioritization algorithms (e.g., rank potential targets given the availability of telescopes $X$, $Y$, and $Z$, with $t$ seconds of available integration time).
\looseness=-1

In order to accommodate the volume of solar system transients needing follow-up, new operational models may need to be considered.  We recommend that relevant agencies consider having facilities with a substantial amount of dedicated target of opportunity time or at least fast-turnaround proposal or Director's Discretionary Time review.  The \textit{Neil Gehrels Swift Observatory} is one potential operational model.  The telescope allows for pointed observational proposals and rapid responses to target of opportunity (ToO) observation requests, but also automatically observes possible gamma-ray events whenever detected.  In our context, the latter would be replaced by targets sourced via observatory alert streams and observation managers, such as near-Earth object or cometary outburst discoveries.  Observatory-level programs could be competed through proposals, but this does not preclude publicly accessible results or short proprietary timescales.
\looseness=-1

The need for additional follow-up facilities to exploit the science coming from the flood of new transients in the LSST was set out by 
\citet{najita2016}. That report not only called for additional wide field multi-object low and moderate resolution spectroscopic instruments, but also called for new developments of infrastructure for time-domain astronomy and computing and data resources to maximize LSST science.
Several Astro2020 Decadal Survey white papers have also discussed the use of astrophysical assets for solar system science \citep{JuanolaParramon2019spacetels, bauer2019, clarke2019}, including next-generation extremely large ground- and space-based telescopes \citep{trilling2019,bowler2019,cartwright2019,hammel2019,neveu2019}. Other authors \citep{kurtz2019, siemiginowska2019} have discussed very similar themes to those of this white paper, such as the need for improved digital infrastructure and shared discovery platforms to support interdisciplinary research and discovery in astrophysics and strengthen the ties between exoplanetary and planetary science.  As observatories and solutions are developed, the complexities of solar system science must be appropriately considered so that NSF and NASA's full scientific interests can excel.
\looseness=-1



\subsection{Computing Infrastructure and Data Management}

A survey of software needs in planetary science conducted in 2019\footnote{\url{https://github.com/MillionConcepts/software-survey-2019}} identified three key ``pain points'' for survey respondents: data access, data handling, and user experience.  Specific suggestions to address issues encountered by users in these areas included the development of wrappers or interfaces to planetary data sources, development of tools for metadata processing or management of locally-mirrored data, handling different types of planetary data, and data exploration, and increased use of shared software standards and libraries.  Looking forward to how future data archives and management systems will be designed, it will be extremely important to keep these issues in mind.
\looseness=-1

The NSF's OIR Lab (NOIRLab) Data Lab\footnote{\url{https://datalab.noao.edu/}} provides a useful model for how future data archive systems, including PDS, can increase data usability, especially for large survey data sets (multiple NEO survey data sets are being archived at PDS).  In addition, the Space Telescope Science Institute (STScI) is experimenting with cloud-based access to its data archive.   Figure~\ref{fig:mast} demonstrates user interest through a pilot program with \textit{Hubble Space Telescope} data sets, and how it quickly exceeded their normal monthly data volumes.  The ease of access coupled with cloud computing resources enables new and efficient ways of studying observatory data sets.  Planetary data should be no exception.
\looseness=-1

\begin{figure}
{\caption{Volume of \textit{Hubble Space Telescope} data accessed through the traditional data archive (MAST) and Amazon Web Services (AWS) during the first six months of public availability (June 2018) in AWS \citep{smith19-aws}.}
\label{fig:mast}}
\end{figure}

Necessary computing infrastructure may include communally available high-performance computing (HPC) facilities, such as those provided by NASA’s High-End Computing Program\footnote{\url{https://hec.nasa.gov/}}.  Connectivity between HPCs and data archives are critical, as it would enable large scale, and possibly prompt, processing of survey data with intensive state-of-the-art methods, such as synthetic tracking to discover objects below the single-exposure detection limit of the data \citep{zhai2020}. 
Different modes of data analysis should be available, and we do not mean to preclude local analysis with personal computer systems.  However, in the era of petabyte-scale data sets, the remote data analysis may be increasingly prevalent as the only feasible method of analyzing enormous data sets given limitations on internet bandwidth as well as local disk storage availability.
\looseness=-1

There has been an increase in the use of commercial clouds and container technology for astronomical data management and analysis \citep{smith2019-wp} and this is being extended to collaborative science analysis platforms. \citet{momcheva2019} discuss a prototype at STScI for HST data, and the NOIRLab Data Lab (discussed above) and LSST Science Platform \citep{juric2019} will be the main access pattern for the majority of LSST end users. These science platforms face challenges with supporting and enabling users to transition their workflows to these new environments, with funding and education needed for users to get the most out of these new platforms and large, rich data sets.
\looseness=-1

Solar system objects present unique challenges for data archive management compared to other areas of astronomy because of their non-sidereal motion, requiring use of orbital elements or ephemeris position predictions, rather than fixed right ascension and declination positions, to link data for individual objects.  The ability to efficiently generate ephemeris predictions for objects in data where they should appear will be extremely useful for precovery and recovery identification (relevant for orbit refinement) and forced photometry (relevant for activity and disruption detection).
As extended objects, comets present additional challenges.  The diversity of their morphologies and spatial scales makes it difficult to develop uniform methods of analysis for all comets, while non-gravitational effects caused by cometary outgassing and the difficulty of accurately measuring nucleus positions of highly active comets can complicate the process of orbit determination, and therefore ephemeris prediction.  
Proper accounting of all of these considerations will be essential in the design of data archive management systems for upcoming petabyte-scale surveys to adequately serve solar system science needs in the next decade.
\looseness=-1

Data archive and management systems should ideally enable a range of investigations, from large-scale analyses of full data sets to small exploratory pilot studies.  Ideally, a data archive system should not simply host data but also provide user-focused tools and interfaces to explore and work with those data.  The ability to perform cross-archive analysis of individual objects or groups of objects, combining NASA's PDS, Mikulski Archive for Space Telescopes (MAST), and Infrared Processing and Analysis Center (IPAC) with LSST and other archives, will also be enormously useful.  Ideally, access to these archives should include a user-friendly searchable front-end and a programmatic access interface that enables data retrieval and analysis in a fully reproducible manner.
\looseness=-1

\subsection{Research \& Analysis Funding\label{section:funding}}

The scientific promise of next-generation large-scale surveys will only be achievable if sufficient financial resources are available to support scientists to perform the anticipated science.  For NASA missions, this support often comes in the form of funding for a core science team, which is then sometimes supplemented by a participating scientist program (PSP), a data analysis program (DAP), or both.  Under such frameworks, a science team is typically tasked with ensuring that the key scientific objectives of a mission are met, while PSPs and DAPs provide opportunities for members of the broader community to also contribute to mission science, enhancing the impact of those missions by contributing valuable outside perspectives as well as additional personnel resources. 
While we expect the NEOSM and Roman Space Telescope survey missions to have at least one of these funding elements, LSST notably does not have any funding dedicated to either preparing for or carrying out solar system science, with NSF and U.S.\ Department of Energy funding instead only supporting construction, survey design, pipeline development,
survey operations, and dark energy science, where supplemental philanthropic funding secured by the LSST Corporation has thus far been insufficient for supporting sustained scientific preparatory work at meaningful FTE levels.
\looseness=-1

In an Astro2020 white paper, \citet{bianco2019_lsst_astro2020} highlighted the fact that 
``minimal and unevenly distributed financial support has created significant challenges for the LSST Science Collaborations, which in turn compromises their planning and preparatory work...leading to potential vulnerabilities.''  Among those vulnerabilities, the authors pointed out that the lack of available funding for a significant number of scientists means that the simultaneous, intensive collaboration needed to prepare for the survey properly cannot occur.  Additionally, given that very few US scientists receive funding specifically to prepare to do LSST science while their international counterparts do include science funding as part of their investment in the project, ``this creates the specific risk that US scientists will be unable to compete with funded international LSST members and lead LSST-based science.''
\looseness=-1

In planetary science in particular, given the high fraction of ``soft money'' planetary scientists (at all career stages) in the field today who rely on grants to support their work (e.g., non-tenured research scientists at universities, or researchers at non-profit institutions), 
the issues raised by \citet{bianco2019_lsst_astro2020} are especially relevant.  
The limited availability of research and analysis (R\&A) funding prevents many such scientists from devoting significant time to 
the project, even though they may have significantly more available research time than university faculty.  If available funding support remains minimal, the scientific impact of the survey for solar system science could be far lower than expected, with the scientific community unable to take full advantage of the massive investment made by funding agencies in construction, development, and operations for the survey.  Given inequities in faculty hiring at research universities \citep[e.g.,][]{zellner2019_astro2020diversity}, the limited funding support for LSST-related research poses an inclusion and diversity concern by depressing participation by women and under-represented minorities, who are disproportionately less likely to hold faculty positions, in what may be the most scientifically impactful survey project in a generation.
\looseness=-1

While the 2023--2032 Planetary Science and Astrobiology Decadal Survey will be completed too late to impact funding for solar system science-related preparatory efforts for LSST, we nonetheless strongly recommend that national funding agencies provide dedicated R\&A funding mechanisms to support scientists carrying out research based on LSST data during and after the survey, perhaps via NASA-like PSPs and DAPs, or R\&A funding programs dedicated to survey-based science.  We also strongly recommend that adequate funding mechanisms for preparatory, in-survey, and post-survey science-related work be budgeted for and implemented for any next-generation large-scale surveys that may be planned in the next decade.
\looseness=-1


\section{Recommendations}
We advocate for several key considerations for large survey data sets in order to address the scientific missions of NSF and NASA, as they relate to planetary science: (1) that funding streams support, encourage, and/or target software and scientific projects that directly analyze LSST or other large data sets, and that support for scientists ideally begins before the start of the survey; (2) observatories and data archive interfaces must consider solar system targets (i.e., moving targets) at the start of system planning; (3) data access is best done through efficient APIs or even cloud-based servers that are well-integrated with solar system target searches; front-ends can use these systems for more traditional analysis tasks; (4) hosted science platforms or cloud-based computing resources will be necessary for large-scale data projects, and can even make small-scale projects more efficient; moreover, increasing the connectivity between archives to enable seamless analyses of cross-archive data sets should inspire innovative scientific studies and discoveries; (5) as astrophysical and solar system discoveries increase with time, and as the connection between discovery engines and follow-up telescopes becomes more efficient (or even automatic), the time pressure on follow-up resources will increase, requiring the reconsideration of traditional observational models.
\looseness=-1


\bibliography{references}

\end{document}